\documentclass[a4paper,11pt]{article}
\pdfoutput=1 
\usepackage{jcappub} 
\usepackage[T1]{fontenc}
\usepackage{graphicx}
\usepackage{amssymb}
\usepackage{amsmath}
\usepackage{color}
\usepackage{cancel}
\usepackage{hyperref}
\hypersetup{urlcolor=blue, colorlinks=true,citecolor=blue}
\usepackage[utf8]{inputenc}
\usepackage{url}
\usepackage[normalem]{ulem}

\def\z{z_{\rm obs}}
\def\m{\mu_{\rm obs}}
\def\nIa{\cancel{\rm Ia}}
\def\Ia{\rm{Ia}}
\def\cos{\boldsymbol{\theta}}
\def\vcos{\boldsymbol{\varphi}}

\title{zBEAMS: A unified solution for supernova cosmology with redshift uncertainties}

\author[1,2]{Ethan Roberts,}
\author[1,4,5]{Michelle Lochner,}
\author[6]{Jos\'e Fonseca,}
\author[1,2,3]{Bruce A. Bassett,}
\author[1]{Pierre-Yves Lablanche,}
\author[1]{Shankar Agarwal}

\affiliation[1]{African Institute for Mathematical Sciences,\\6 Melrose Road, Muizenberg, 7945, Cape Town, South Africa}
\affiliation[2]{Department of Maths and Applied Maths, University of Cape Town,\\Cape Town, South Africa}
\affiliation[3]{South African Astronomical Observatory,\\Observatory, Cape Town, 7925, South Africa}
\affiliation[4]{SKA South Africa, 3rd Floor, The Park, Park Road,\\Pinelands, 7405, South Africa}
\affiliation[5]{Department of Physics and Astronomy, University College London,\\Gower Street, London WC1E 6BT, UK}
\affiliation[6]{Department of Physics \& Astronomy, University of the Western Cape,\\Cape Town 7535, South Africa}

\emailAdd{rbreth001@myuct.ac.za}

\abstract{Supernova cosmology without spectra will be an important component of future surveys such as LSST. This lack of supernova spectra results in uncertainty in the redshifts which, if ignored, leads to significantly biased estimates of cosmological parameters. Here we present a hierarchical Bayesian formalism -- zBEAMS -- that addresses this problem by marginalising over the unknown or uncertain supernova redshifts to produce unbiased cosmological estimates that are competitive with supernova data with spectroscopically confirmed redshifts. zBEAMS provides a unified treatment of both photometric redshifts and host galaxy misidentification (occurring due to chance galaxy alignments or faint hosts), effectively correcting the inevitable contamination in the Hubble diagram. Like its predecessor BEAMS, our formalism also takes care of non-Ia supernova contamination by marginalising over the unknown supernova type. We illustrate this technique with simulations of supernovae with photometric redshifts and host galaxy misidentification. A novel feature of the photometric redshift case is the important role played by the redshift distribution of the supernovae.}

\begin{document}
\maketitle
\flushbottom

\section{Introduction}

Studies of Type Ia supernovae led to the dark energy breakthrough and modern concordance cosmology, but one can argue that they have been supplanted by Baryon Acoustic Oscillations \cite{BAO1, BAO2} as the most precise way of constraining cosmology today. To be competitive in the era of LSST \cite{LSST2009}, EUCLID and the SKA, supernova cosmology faces several big challenges. One is that better control of systematics is required and a number of sophisticated approaches are being developed to improve the control of systematics (e.g. \cite{Mandel2009, Rubin2015, kesslerBEAMS, BAHAMAS, Ma, Jennings, Betoule2014, Chambers2016}). Another critical problem is that next-generation supernova surveys will be severely spectroscopy limited: LSST will deliver over $10^5$ Type Ia Supernova (SNIa) candidates with photometric lightcurves only. The lack of spectroscopy introduces a number of challenges. First, the true identity of any candidate without spectroscopic follow-up is ambiguous - photometric colours only provide a probability for an object to be a SNIa, as opposed to a Type Ibc or II supernova or other transient \cite{psnid}. Secondly, the precise redshifts of the supernovae are unknown. Photometric redshifts are fairly good if the candidates are {\em known} to be SNIa, yielding RMS errors of $\sigma_z \sim 0.04(1+z)$, depending on exact assumptions  \cite{LSST2009, Kessler2010, Wang2015}. The problem is that we are exactly in the case where we are not sure whether each candidate is a SNIa or not, and the photometric redshift error is much larger if the object is not a SNIa \cite{Moller2016}, precisely because they are not standard candles.

A promising approach that dates back to the SDSS II supernova survey \cite{Hlozekbeams, Campbell, Olmstead2013}, is to obtain spectroscopic redshifts for the host galaxies of the supernova candidates and use this as a proxy for the supernova redshift.  This will be particularly attractive in the era of big redshift surveys such as 4MOST, SKA and Euclid, where huge numbers of galaxy redshifts will be known. This has the potential to help remove biases \cite{Olmstead2013} and yield improved constraints \cite{Yuan}.

\begin{figure}[tbp]
\includegraphics[scale=0.43]{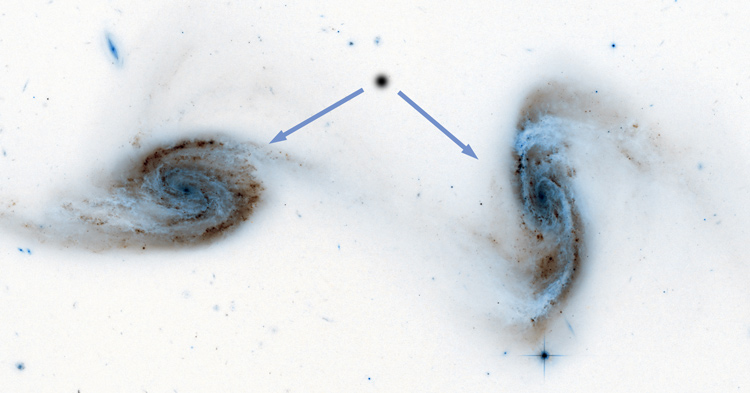}
\centering
\caption{Schematic figure illustrating the source of redshift contamination: the host galaxy of a supernova may be ambiguous even if the redshifts of the galaxies nearby are known. Here we assume that there is an estimate of the probability, denoted $P(\gamma)$, that the supernova belongs to each galaxy (where each galaxy is indexed by $\gamma$).}
\label{fig:hosts}
\end{figure}

However, even this approach has a serious problem: identifying the host galaxy is also not unambiguous. The supernova can appear to lie in between two or more galaxies or may live in a host that is too faint to be detectable (``hostless'') (see figure~\ref{fig:hosts}). In general we therefore can assume that instead we have probabilities for the supernova to belong to each of the nearby galaxies on the sky or to be hostless. Current matching algorithms can accurately match the correct host galaxy about $91\%$ of the time when applied to data, potentially increasing to $97\%$ by using machine learning techniques \cite{Gupta2016}. However, even a $3\%$ contamination may cause significant biases on cosmological parameter inference and must be dealt with.

Our goal with this paper is to develop a single formalism that solves this set of problems, simultaneously handling both the contamination from non-Ia supernovae and contamination from incorrect host identification and redshift assignment. The formalism we desire will be rigorous without resorting to cuts, which are statistically suboptimal.  

In laying out the solution we will develop intuition by building systematically to the complete solution. In section~\ref{sec:standardcosmo} we review inference using standard cosmology. In section~\ref{sec:Single_SNIa} we consider the case of a single, known Type Ia supernova with a photometric redshift derived from the supernova lightcurve. In section~\ref{sec:Host_galaxy} we extend the analysis to include host galaxy redshifts. In section~\ref{sec:NonIa} we review the BEAMS formalism to handle contamination from non-Ia. In section~\ref{sec:zBEAMS} we present the combined formalism to handle all contamination and finally in section~\ref{sec:simulations} we present an illustrative set of simulations demonstrating our method.

\section{Standard Supernova Cosmology Inference} \label{sec:standardcosmo}

Traditionally, supernova cosmological analysis proceeds with a sample of spectroscopically confirmed type Ia supernovae with well-measured redshifts. The goal is to determine the posterior distribution, $P(\cos|\{D_i\})$, over the cosmological parameters, $\cos$, given the dataset which we denote $\{D_i\}$.\footnote{In much of our discussion we will use $D$ to denote a single supernova so for clarity we use $\{D_i\}$ to make explicit the case where we are considering multiple supernovae.} Here, the data are the redshift, $z_i$, and distance modulus, $\mu_i$, for each supernova. The distance modulus for a type Ia supernova is usually estimated from its observed light curve using the SALT2 model \cite{Guy2007}.

The distance modulus is defined as:

\begin{equation}
 \mu(z) = m - M = 5 \text{log}_{10}\left(\frac{d_L}{1Mpc}\right) + 25,
\end{equation}
where $m$ is the apparent magnitude of the object, $M$ is the absolute magnitude of the object and $d_L$ is the luminosity distance to the object in Mpc. In a $\Lambda$CDM universe, the luminosity distance is related to the cosmological parameters by:
\begin{equation}
 d_L (z) = \frac{c(1+z)}{H_0 \sqrt{-\Omega_k}} \text{sin}\left( H_0 \sqrt{-\Omega_k} \int \frac{dz'}{H(z')} \right),
 \end{equation}
where 
\begin{equation}
 H(z) = H_0 \bigg(\Omega_m (1+z)^3 + \Omega_{\rm DE} (1+z)^{3(1+w)} + \Omega_k(1+z)^2 \bigg)^{1/2},
\end{equation}
and $H_0$ is the Hubble constant, $\Omega_m$ is the energy density of matter, $\Omega_{\rm DE}$ is the energy density of the dark energy, $w$ is the dark energy equation of state where $w = -1$ corresponds to $\Lambda$, the cosmological constant. Finally $\Omega_k$ is the curvature parameter. Collectively, we refer to these cosmological parameters as $\cos$. 

The posterior probability distribution over the cosmological parameters is then given by Bayes' theorem:

\begin{equation}
P(\cos|\{D_i\}) = \frac{P(\{D_i\}|\cos) P(\cos)}{P(\{D_i\})},
\end{equation}
where $P(\{D_i\}|\cos)$ is the likelihood, $P(\cos)$ is the prior and $P(\{D_i\})$ is called the Bayesian evidence. Since we are not interested in model selection $P(\{D_i\})$ is irrelevant for parameter inference and we drop it in all subsequent analysis. 

In the case of uncorrelated Gaussian errors, $\sigma^i_\mu$, on the estimated distance moduli $\m^i$, the likelihood over the $N$ supernovae is: 

\begin{equation}
P(\{D_i\}|\cos) = \prod_i^N \,\frac{1}{\sqrt{2\pi {\sigma^i_\mu}^2}} \rm{exp} \bigg(-\frac{(\m^i - \mu(z^i, \cos))^2}{2{\sigma^i_\mu}^2} \bigg)\,.
\end{equation}

A standard cosmological analysis would then proceed by applying a numerical sampling method such as MCMC \cite{Metropolis1953, Hastings1970} to determine the full posterior of the cosmological parameters. Care needs to be taken around the parameters involved in estimating the distance modulus, such as considered in \cite{March2011}. While such caveats can be handled easily by introducing new latent parameters, for simplicity we omit these parameters and assume the distance moduli can be measured directly since it is not core to the problem we are addressing.

\section{Inference in the presence of redshift uncertainties} \label{sec:Single_SNIa}

To achieve our goal of dealing with the unknown redshifts and contamination we will need two pieces of Bayesian technology in addition to Bayes theorem: marginalisation and the product rule, which we briefly review here. The Product Rule states that, for any sets of parameters $\cos,\vcos$ and ${\bf C}$, we have:
\begin{equation}
P(\cos, \vcos | {\bf C}) = P(\cos | \vcos, {\bf C}) P(\vcos |{\bf C})\,.
\end{equation}
The second piece of technology is marginalisation over nuisance parameters, $\vcos$: 
\begin{equation}
P(\cos |{\bf C}) = \int P(\cos, \vcos | {\bf  C}) ~d \vcos \,.
\end{equation}
The combination of these symbol manipulation techniques will be useful to deal with latent parameters; such as the true (unknown) redshift, $z$, and type $\tau$, of the supernova. 

As a step to the fully general case, let us consider a single, known Type Ia supernova with an uncertain redshift, such as that derived from the supernova lightcurve. Very small spectroscopic redshift uncertainties, $\delta z$, have usually been dealt with by converting them into an additional error in the distance modulus, $\delta \mu$, by assuming a model $\mu(z)$ for the conversion, and adding the result in quadrature with the usual $\mu$ error. As we will show in detail later this is not statistically correct (for one since we don't know the correct distance modulus describing our universe) and fails badly for typical photometric redshift errors. 

To proceed rigorously we instead start by using Bayes theorem for the posterior for $\cos$ given some  data, $D$, from a single supernova. We then expand the arguments of the posterior to include the true redshift ($z)$ and distance modulus ($\mu$) of the supernova as latent (i.e. nuisance) parameters that we marginalise over, since we don't know their true values:  
\begin{eqnarray}
P(\cos | D) &\propto&  P(D | \cos) P(\cos) \,,\\
&\propto& \int P(D, z, \mu | \cos) P(\cos) \, dz \, d\mu \,.
\end{eqnarray}
We repeatedly apply the product rule to rewrite this multi-dimensional integral as:
\begin{eqnarray}
P(\cos | D) &\propto& \int P(D | z, \mu, \cos) P(z, \mu | \cos) P(\cos) \, dz \, d\mu\,, \\
&\propto& \int P(D | z, \mu) P(\mu | z, \cos) P( z |  \cos) P(\cos) \, dz \, d\mu \,.
\label{eq:zprod}
\end{eqnarray}
We now make the simplifying assumption that the distribution of the true redshift is independent of the cosmological parameters\footnote{A more accurate assumption would be to posit that it depends on $\cos$ only via the volume of spacetime. This is expected to be a weak dependence for the currently allowed range for $\cos$.} and note that since $\mu$ is assumed to be a deterministic function of $z$ and $\cos$ the distribution $P(\mu | z, \cos)$ is a delta function\footnote{This is only exactly true in the background FLRW model and is not true if one allows for effects such as gravitational lensing, but we will ignore such perturbative effects here as is typical in supernova studies.}, which allows us to eliminate the $\mu$-integral\footnote{Using the standard identity $\int f(x) \delta(x-x_0) dx = f(x_0)$}: 
\begin{eqnarray}
P(\cos | D) &\propto& \int P(D | z, \mu) \, \delta(\mu - \mu(z, \cos)) P(z) P(\cos) \, dz \, d\mu\,, \\
&\propto& \int P\left(D | z, \mu(z, \cos)\right) \, P(z) \, P(\cos) \, dz \,.
\end{eqnarray}
This is the expression we were seeking. It expresses the posterior as a marginalisation over the unknown supernova redshift. What is the data $D$? In this case let us assume that we have extracted both an estimate of the redshift and distance modulus, ($\z,\m$), for the supernova from its lightcurve. We now have:
\begin{eqnarray}
P(\cos | D) &\propto& P(\cos) \int_0^{\infty} P\left(\z, \m | z, \mu(z, \cos)\right) \, P(z) \, dz \,.
\label{eq:zpostn}
\end{eqnarray}
Assuming that the estimates $\z$ and $\m$ are independent, we can simplify this to: 
\begin{eqnarray}
P(\cos | D)  &\propto& P(\cos) \int_0^{\infty} P\left(\z | z \right) P\left(\m |\, \mu(z, \cos)\right) \, P(z) \, dz\,.
\label{eq:zpost1}
\end{eqnarray}

An important novel role in this analysis is played by the redshift prior $P(z)$. This has no parallel in the usual supernova analysis where the redshift is known spectroscopically. In the case of the uncertain supernova redshift we must consider Eddington bias: a supernova discovered with a 4m telescope with a redshift estimate of $\z = 0.75$, is much more likely to lie at a {\em true} redshift of $z=0.6$ than at $z = 0.9$.

This insight, encoded in the prior $P(z)$, must be included in our inference analysis. Gull \cite{gull} was perhaps the first to highlight the importance of the data prior, showing that the use of the wrong prior (i.e. not the true distribution from which the data was drawn) leads to a bias. 

Including this is crucial in any practical application since we are not sure about the true redshift prior, $P(z)$, which will depend both on the cosmological volume (which depends on the cosmological parameters) and the rates for each type of supernova. We can handle this uncertainty by introducing hyperparameters, $\varphi$, into the prior, $P(z, \varphi)$ which can then be fit for or marginalised over. Indeed, one can turn this around and instead marginalise over the cosmological parameters and supernova redshifts to produce a posterior for the rates of the different supernova types as functions of redshift\footnote{Numerical marginalisation is achieved simply by selecting and histogramming the parameters of interest from the MCMC chain.}. Hence this ``annoyance'' which cannot be handled by the standard formalism becomes a feature in our analysis. 

While this will be important for realistic simulations and analysis of real data it is a conceptually straightforward Bayesian extension and we do not consider it further here.   

We now consider two special limiting cases to gain some intuition: (i) Spectroscopic galaxy redshifts but unknown host identity, and (ii) photometric supernova redshifts alone. 

\subsection{Spectroscopic galaxy redshifts but unknown host galaxy identity} \label{sec:Host_galaxy}

In this subsection we answer the question: what happens if we have spectroscopic redshifts for all galaxies but are not sure which galaxy is the true host? 

Since we take our data to be coming from the supernova lightcurve only, we can treat spectroscopic galaxy redshifts as a prior $P(z)$ on the SN redshift. Since we don't know the true host in general we write this prior as a marginalisation over the potential host galaxies, which we index by $\gamma$: 
\begin{eqnarray} \label{eq:priorz}
P(z) &=& \sum_\gamma P(z | \gamma) P(\gamma)\,.
\end{eqnarray}
In the case of figure~\ref{fig:hosts} this sum would likely just consist of just three terms: the two nearby galaxies and a third invisible galaxy which might be too faint to be detected in the image. 

In eq.~\eqref{eq:priorz} $P(\gamma)$ is the probability that the supernova is hosted by the $\gamma^{\rm th}$ galaxy\footnote{Estimated for example using the projected distance from the centre of each galaxy of some other measure.}, and $P(z | \gamma)$ is the corresponding redshift distribution for the $\gamma^{\rm th}$ galaxy. Then we have:
\begin{eqnarray}
P(\cos | D) &\propto& P(\cos) \int dz \sum_\gamma P(z | \gamma) P(\gamma) P\left(\z, \m | z, \mu(z, \cos)\right)
\label{eq:galhostinfo}
\end{eqnarray}
In the limit where the galaxy redshifts are known spectroscopically to high precision we can approximate the host redshift distribution as a delta function: $P(z | \gamma) = \delta(z-z_\gamma)$ where $z_\gamma$ is the spectroscopically determined redshift of the $\gamma^{\rm th}$ galaxy. Then we can perform the marginalisation analytically, yielding:
\begin{eqnarray}
P(\cos | D) &\propto& P(\cos) \sum_\gamma P(\gamma) P\left(\z, \m | z_\gamma, \mu(z_\gamma, \cos)\right)\,.
\label{eq:spectroz}
\end{eqnarray}
In other words, the final posterior is a mixture model that simply sums all the posteriors arising from assuming the supernova belongs to each of the potential host galaxies, weighted by the probability $P(\gamma)$ that the supernova belongs to each host. This is an intuitively pleasing and simple result. 

\subsection{Photometric redshifts} \label{sec:photoz}

A second limiting subcase that will give some useful intuition is the case where we have no spectroscopic host information but instead only have a photometric redshift estimate from the supernova itself or from the host galaxy in the case where the host is unambiguous, i.e. we have an estimate for $P\left(\z | z \right)$. 

For simplicity assume that the resulting photometric redshift distribution is Gaussian\footnote{The generalisation to an arbitrary distribution is in principle simple since we perform the marginalisation numerically. The formalism remains unchanged with the new photometric redshift distribution.}. If we assume $\z$ and $\m$ are correlated, for example if they both come from the lightcurve, then eq.~\eqref{eq:zpostn} becomes:

\begin{equation}
P(\cos | D) \propto P(\cos) \int \frac{1}{2\pi \sqrt{det|C|}} \exp\left(-\frac{1}{2} \Delta^T C^{-1} \Delta \right) P(z) dz\,,
\end{equation}
where $\Delta = \begin{pmatrix}
\mu_{obs}-\mu \\
z_{obs}-z
\end{pmatrix}$ and $C = \begin{pmatrix}
\sigma_{\mu}^2 & \sigma_{\mu z}\\
\sigma_{\mu z} & \sigma_z^2
\end{pmatrix}$ is the covariance matrix.

Assuming independent $\z$ and $\m$ estimates for simplicity, eq.~\eqref{eq:zpostn}, reduces to:
\begin{equation}
P(\cos | D) \propto P(\cos) \int \frac{1}{2\pi \sigma_z \sigma_\mu} \exp\left(- \frac{(\z - z)^2}{2\sigma_z^2} \right) \exp\left(-\frac{(\m - \mu(z, \cos))^2}{2\sigma_\mu^2} \right) P(z) dz\,.
\label{eq:anights}
\end{equation}

In our simulations we numerically perform the marginalisation in eq.~\eqref{eq:anights} without considering correlations between $\z$ and $\m$, which is not generally true. However such correlations can be included in a straight forward way via modeling the covariance function, potentially with hyperparameters which can also be marginalised over. 

Examining eq.~\eqref{eq:anights} one might be tempted to Taylor expand $\mu(z,\cos)$ around $\z$. Taking only the linear term in $\Delta z \equiv (z-\z)$ one can, assuming a Gaussian prior $P(z)$ with mean $\bar{z}$ and standard deviation $\sigma_p$, perform the redshift marginalisation analytically \footnote{Note that one cannot simply assume an improper uniform prior since this leads to biases with real data while assuming a proper top-hat prior is no longer analytically integrable.}, giving:
\begin{equation}
P(\cos|D) \propto P(\cos)\int \exp \left( -\frac{(\z-z)^2}{2\sigma_z^2} -\frac{(\tilde{\mu}_{\rm obs}-\mu'z)^2}{2\sigma_\mu^2} \right) \exp \left( -\frac{(z-\bar{z})^2}{2\sigma_p^2} \right) dz\,,
\end{equation}
where $\tilde{\mu}_{\rm obs} = \m - \mu(\z) + \mu'z_{\rm obs}$. Completing the square and performing the integral gives:
\begin{equation}
P(\cos|D) \propto  \frac{P(\cos)}{C^{1/2}} \exp\left(-\frac{1}{2C} \left[ \sigma_\mu^2 (\z-\bar{z})^2 + \sigma_p^2(\m-\mu(\z))^2 + \sigma_z^2(\tilde{\mu}_{\rm obs}-\mu'\bar{z})^2 \right]\right)\,,
\label{eq:photozfinal1}
\end{equation}
where $C$ is equal to :
\begin{equation}
C = \sigma_\mu^2(\sigma_p^2+\sigma_z^2)+\mu'^2\sigma_z^2\sigma_p^2\,.
\end{equation}

This has a  simple interpretation: the $z$-error is converted into a $\mu$ error by using $\mu^{\prime}$, the instantaneous slope of the $\mu(z,\cos)$ curve, which is then added in quadrature with the existing $\mu$ error, $\sigma_{\mu}^2$; all modulated by the redshift prior width $\sigma_p$. Since we are assuming a Gaussian likelihood this is the same result as for the corresponding Fisher matrix analysis, see \cite{doublefisher}. Note that to do this self-consistently one must change the $\mu$ error for each supernova at every point in the MCMC chain (since changing $\cos$ changes $\mu^{\prime}$). 

On the surface this offers a big computational simplification. For a large dataset of $N$ supernovae and $m$ cosmological parameters, eq.~\eqref{eq:photozfinal1} only requires one to fit for the $m$ parameters rather than all $N+m$ parameters. Unfortunately in most cases eq.~\eqref{eq:photozfinal1} is not sufficient for several reasons:
\begin{itemize}
\item It is only applicable when $P(\z | z)$ is sufficiently narrow, i.e. $\sigma_\mu^2  \gg \sigma_z^2 \mu^{\prime 2}$. But what does this mean in practise?  

\item As alluded to previously we can only perform the integral in eq.~\eqref{eq:anights} analytically for special priors such as an improper uniform ($-\infty < z < \infty$) or Gaussian prior.\footnote{This is integrable for a prior $P(z) \propto z e^{-\beta z}$ but this prior doesn't have enough freedom in general for representing supernova rates in astronomy.} These are not appropriate for astronomical surveys and hence assuming one of them will give biased results in general \cite{gull}. 

\item In the general case we do not know the type of the supernova for sure. This implies that the photometric redshift for the supernova may be very non-Gaussian in general. This in turn implies that we cannot analytically integrate eq.~\eqref{eq:anights} even if we assume a Gaussian redshift prior. 
\end{itemize}
In the simulations below we find that this approach is biased. For all these reasons we strongly recommend  numerically marginalising over all redshifts via MCMC or suitable sampling technique as we present in subsection~\ref{sec:photoz}.

\section{Contamination from non-Ia supernovae} \label{sec:NonIa}

In the previous section we derived the posterior in the presence of redshift contamination but assumed that we knew the object was a SNIa. Unfortunately contamination of supernova types is also inevitable for photometric surveys. This has been addressed by the BEAMS (Bayesian Estimation Applied to Multiple Species) formalism for the case of spectroscopic redshifts in various papers \cite{Kunz2007, Knights2012, Rubin2015, Hlozekbeams}. Here we highlight the key results using the hierarchical Bayesian approach we have used so far in this paper for redshift contamination as a warm-up to the case with both types of contamination.  

In this case we can assume we know the true redshift of the supernova but are not sure of its type, labeled by a discrete variable $\tau$, which we here allow to take two values: $\tau = $ Ia and $\tau = \nIa$, the latter corresponding to non-Ia objects.\footnote{If there were more than one class of object that gave useful information about $\cos$ we would need to subdivide the non-Ias into more classes; e.g. if the $\cos$ represented information about star formation rates rather than cosmology.}

Previously we marginalised over the unknown latent variables $z$ and $\mu$. Now instead we marginalise over $\tau$, $\mu$ and $z$. Following the same approach taken in eq.~\eqref{eq:zprod} and thereafter, i.e., using Bayes theorem and the Product Rule to marginalise over the latent variables $\tau$, $\mu$ and $z$, one has

\begin{eqnarray}
P(\cos | D) &\propto&  P(D | \cos) P(\cos)\,, \\
&\propto& \int  P(D|z,\mu,\tau,\cos)P(z,\mu|\tau,\cos)P(\tau)P(\cos) \, d\tau \, dz \, d\mu\,,\\
&\propto& \int P(D|z,\mu,\tau,\cos)P(\mu|z,\tau,\cos)P(z | \tau, \cos)P(\tau)P(\cos)\, d\tau \, dz \, d\mu\,.
\end{eqnarray}
Since we assume we know the redshift of the supernova, $z_*$, perfectly, we can write $P(z | \tau, \cos)$ as a delta function $\delta(z-z_*)$ which allows us to do the $z$-integral, yielding:
\begin{eqnarray} \label{eq:beams}
P(\cos | D) & \propto & \int  P(D|z_*,\mu,\tau,\cos)P(\mu|z_*,\tau,\cos)P(\tau)P(\cos)\, d\tau \, d\mu\,.
\end{eqnarray}
As before, $\mu$ is a deterministic function of $z$, $\tau$ and $\cos$ and hence  $P(\mu|z_*,\tau,\cos)$ is also a delta-function, which collapses the $\mu$-integral. We write $\mu_{\Ia}$ for the expression $\mu(z, \cos, \tau = \Ia)$ and $\mu_{\nIa}$ for $\mu(z, \cos, \tau = \nIa)$. The $\tau$ `integral' is actually just a sum over the two supernova classes. Here, $P(\tau)$ is the supernova type probability.\footnote{This probability is estimated directly from the lightcurve using templates, inference or machine learning \cite{psnid, challenge}.} For notational simplicity we write the probability of being a type Ia supernova as $P_{\Ia}$ (as for $P(\tau = \Ia)$) and the probability of not being a type Ia supernova as $P_{\nIa} = 1 - P_{\Ia}$. 

Then eq.~\eqref{eq:beams} reduces to the usual BEAMS result:
\begin{equation} \label{eq:sum}
P(\cos|D) \propto  P(\cos) \left[ P_{\Ia}\,P(D|z_*,\mu_{\Ia}(z), \tau=\Ia,\cos)  + (1-P_{\Ia}) \,P(D|z_*,\mu_{\nIa}(z),\tau=\nIa,\cos) \right]\,.
\end{equation}
For a Gaussian likelihood with observed distance modulus $\m$, we can write:
\begin{equation}
P(D|z_*,\mu_{\Ia}(z), \tau=\Ia,\cos) = \frac{1}{\sqrt{2\pi}\sigma_{\Ia}}\exp\left( \frac{-(\m - \mu_{\Ia}(z_*,\cos))^2}{2\sigma_{\Ia}^2} \right) \,.
\end{equation}
What about the non-Ia term $\mu_{\nIa}$? Following \cite{Hlozekbeams} and \cite{Kunz2007} one can take $\mu_{\nIa} = \mu_{\Ia} + b(z,\varphi)$ where $b(z,\varphi)$ allows both for the redshift evolution of the non-Ia population and the difference in mean intrinsic luminosity compared with SNIa, parameterised by some unknown hyperparameters $\varphi$ which are fit simultaneously with all the other parameters. This allows one to learn about the non-Ia populations. For cosmology, the non-Ia population is essentially uninformative and hence one can simply take the corresponding $\sigma_{\nIa}$ to be large in actual analysis.

\section{The general case: both type and redshift contamination} \label{sec:zBEAMS}

We are now ready to consider the general case with both type and redshift contamination. Hence we assume that the type $\tau$ of the supernova is unknown, and that we either have photometric redshift estimate for the supernova or redshift information (either photometric or spectroscopic) of potential hosts galaxies.

As before we will treat the host galaxy information as a prior, $P(z)$, on the SN redshift (see eq.~\eqref{eq:priorz})\footnote{The supernova rates are now a much less important factor in the prior since the redshift priors arising from the galaxies should be very peaked in comparison.}: 
\begin{equation}
P(z) = \sum_\gamma P(z | \gamma) P(\gamma)\,.
\end{equation}
These need not be spectroscopic redshifts and $P(z | \gamma)$ may have significant spread.

As before we begin from Bayes rule, introduce and marginalise over latent variables for the unknown true variables, $z, \mu, \tau$, and then repeatedly apply the product rule:
\begin{eqnarray}
P(\cos | D) 
&\propto& \int P(D, \tau, z, \mu | \cos) P(\cos)\, d\tau \, dz \, d\mu \,,\\
&\propto& \int P(D | \tau, z, \mu, \cos) P(\tau, z, \mu | \cos) P(\cos) \, d\tau \, dz \, d\mu\,, \\
&\propto& \int P(D | \tau, z, \mu) P(\mu | \tau, z, \cos) P(\tau, z |  \cos) P(\cos) \, d\tau \, dz \, d\mu\,,\\
&\propto& \int P(D | \tau, z, \mu) P(\mu | \tau, z, \cos) P(\tau | z, \cos) P(z | \cos) P(\cos) \, d\tau \, dz \, d\mu \,,\\
&\propto& \int P(D | \tau, z, \mu) \delta(\mu - \mu(\tau, z, \cos)) P(\tau | z) P(z) P(\cos) \, d\tau \, dz \, d\mu \,,\\
&\propto& \sum_{\tau} \int P\left(D | \tau, z, \mu(\tau, z, \cos)\right) P(\tau | z) P(z) P(\cos) \, dz\,,
\end{eqnarray}
where the last step accounts for the fact that $\tau$ is a discrete (categorical) variable. We can now substitute for the prior $P(z)$ expressed as the sum over the potential host galaxies, eq.~\eqref{eq:priorz}, which gives: 
\begin{eqnarray} \label{eq:zBEAMS_1}
P(\cos | D)  &\propto& P(\cos) \int dz \, \sum_\tau P(\tau | z) P\left(D | \tau, z, \mu(\tau, z, \cos)\right)   \sum_\gamma P(z | \gamma) P(\gamma)\,.
\end{eqnarray}
Eq.~\eqref{eq:zBEAMS_1} is our main result for the posterior arising from a single supernova. It handles both Type and redshift-host contamination. Note that because the true redshift of the supernova is unknown we can allow for redshift-dependence of the type probabilities, $P(\tau | z)$, whereas in previous BEAMS analysis this was taken as a constant. 

To compute the full posterior over $N$ supernovae with collective data $\{D_i\}$, we now have $2N$ nuisance parameters corresponding to the type and redshift, $(\tau, z)$ for each supernova, in addition to the cosmological parameters $\cos$. We can then either assume the supernovae are independent, and simply multiply the posteriors to get:
\begin{equation}
P(\cos | \{D_i\})  \propto P(\cos) \prod_i^N \,\left[ \int dz \, \sum_{\tau_i} P(\tau_i | z_i) P\left(D_i | \tau_i, z_i, \mu(\tau_i, z_i, \cos)\right)  \sum_{\gamma_i} P(z_i | \gamma_i) P(\gamma_i) \, \right]\,,
\label{eq:fullposterior}
\end{equation}
or allow for correlations between the supernovae as in \cite{Knights2012}.

In practise the marginalisation over redshift required by eq.~\eqref{eq:fullposterior} can be achieved efficiently through MCMC by allowing the redshift of each supernova to be a free nuisance parameter that is varied along with the cosmological parameters $\cos$. If the supernovae are correlated then we must also introduce the type of each supernova to include as a parameter. This does not significantly alter the MCMC analysis \cite{Knights2012}.

At this point one can again ask what the data, $D_i$, is for each supernova? At the most basic -- and correct -- level this would be the lightcurve measurements in various colour bands as a function of time.  However, a convenient simplification is to consider $D_i = (\z, \m)$, {\em assuming} that the object is a SNIa. In general this would be inappropriate if more than one type of supernova contained useful information about the cosmological parameters $\cos$. However, since in the non-Ia case the derived redshift and distance modulus give almost no useful cosmological information, one can simply take $P(\z | z, \tau =  \nIa)$ to be a wide uniform distribution or very wide Gaussian and use $\z,\m$ derived assuming the object is a SNIa \cite{Kunz2007, Rubin2015}. Then the fact that one is using the ``wrong'' values for $\z,\m$ has no impact. We note that obviously this is only an issue when the type of the supernova is unknown. 

\subsection{Dealing with Selection Bias} \label{malmquist}

Our eq. (\ref{eq:fullposterior}) is not completely general of course. It neglects the fact that real surveys will have censored data due to selection effects, also known as Malmquist bias: the faint end of the population will not be detected, especially at high redshift. If untreated this will bias the cosmological results.  One approach is to apply a Malmquist bias correction directly to the data before undertaking the analysis, as done in the analysis of the SDSS-II data in \cite{Hlozekbeams}. This is unsatisfactory from a Bayesian point of view however. Another approach, following e.g. \cite{Rubin2015}, is to consider the full population of supernovae, only a subset of which are actually detected by any given SN survey. Introducing a latent label $\epsilon$ which is unity if the SN is detected by the survey and zero otherwise, we can then treat $\epsilon$ as we have the other latent variables and marginalise over it, which will lead to the appearance of the probability distribution $P(\mbox{detect} | D)$ which will depend explicitly on the telescope and survey strategy. In addition, we will need to marginalise over the unknown number of undetected supernovae.  

Within our current approach of treating the data to be the distance modulus and redshift, we could simply truncate the likelihood at a given threshold value of $\mu$. This has been implemented in the context of a hierarchical Bayesian model like the one used here by Sereno \cite{Sereno2016}. However, a simple censoring of distance moduli will not capture the true complexities of selection effects in real surveys. A better approach will be to go back to the lightcurve flux measurements directly but this in turn will depend on the specific lightcurve fitter one uses; see \cite{Rubin2015} for an analysis of Malmquist bias in the context of the SALT II lightcurve fitter and e.g. \cite{Mackay2003} section~3.1 for a pedagogical introduction to dealing with selection effects within Bayesian statistics. In such studies a key role is played by demanding that the censored likelihood is still normalised. 

We will ignore selection bias in our simulations since to do so correctly would require working at the level of light curves, which we have avoided for simplicity, and the standard approach still fails dramatically even in this idealised case. We now move to test the zBEAMS formalism in different cases.

\section{Simulations}\label{sec:simulations}

Here we perform an illustrative set of catalogue simulations to show how zBEAMS, as described in sections~\ref{sec:Single_SNIa} to \ref{sec:zBEAMS}, recovers the correct cosmological parameters by marginalising over unknown supernova types and redshifts. We consider the two cases corresponding to sections~\ref{sec:zBEAMS} and \ref{sec:photoz}: one where we have spectroscopic redshift estimates from potential host galaxies, hereafter referred to as the {\it spectroscopic case}, and one where we have photometric measurements only, hereafter referred to as the {\it photometric case}. Conceptually, it is trivial to combine these cases when dealing with a dataset with mixed spectroscopic and photometric measurements, however we keep the two cases distinct. 

In both cases for simplicity we assume that all objects are detected, no matter how faint. We demonstrate that while the standard likelihood, ignoring redshift uncertainty, results in biased cosmological parameter estimates even in this case, zBEAMS correctly marginalises over type and redshift uncertainties, recovering the fiducial cosmology. 

For all simulations, we assume a flat $\Lambda$CDM universe with a fiducial cosmology given by the latest results from the Planck collaboration \cite{Planck2015}, i.e., $H_0=67.74 ~\rm{km/s/Mpc}$, $\Omega_m=0.31$ and $w=-1$. We perform inference over the parameters using Markov Chain Monte Carlo (MCMC) methods, specifically we use the Metropolis-Hastings \cite{Metropolis1953, Hastings1970} algorithm for low-dimensional sampling and block Metropolis-Hastings when numerically marginalising over redshift in the photometric case. Detailed descriptions of both cases follow below.

\subsection{Spectroscopic Case}
For the spectroscopic case, we simulate 1000 SNe from a uniform distribution across a redshift range of $z=0.015-1$. We create two datasets, one pure Ia dataset without any redshift errors for reference, herein referred to as the unbiased dataset, and one with both host galaxy redshift and non-Ia contamination, herein referred to as the biased dataset.     

\begin{figure}[tbp]
\centering
\includegraphics[width=14cm]{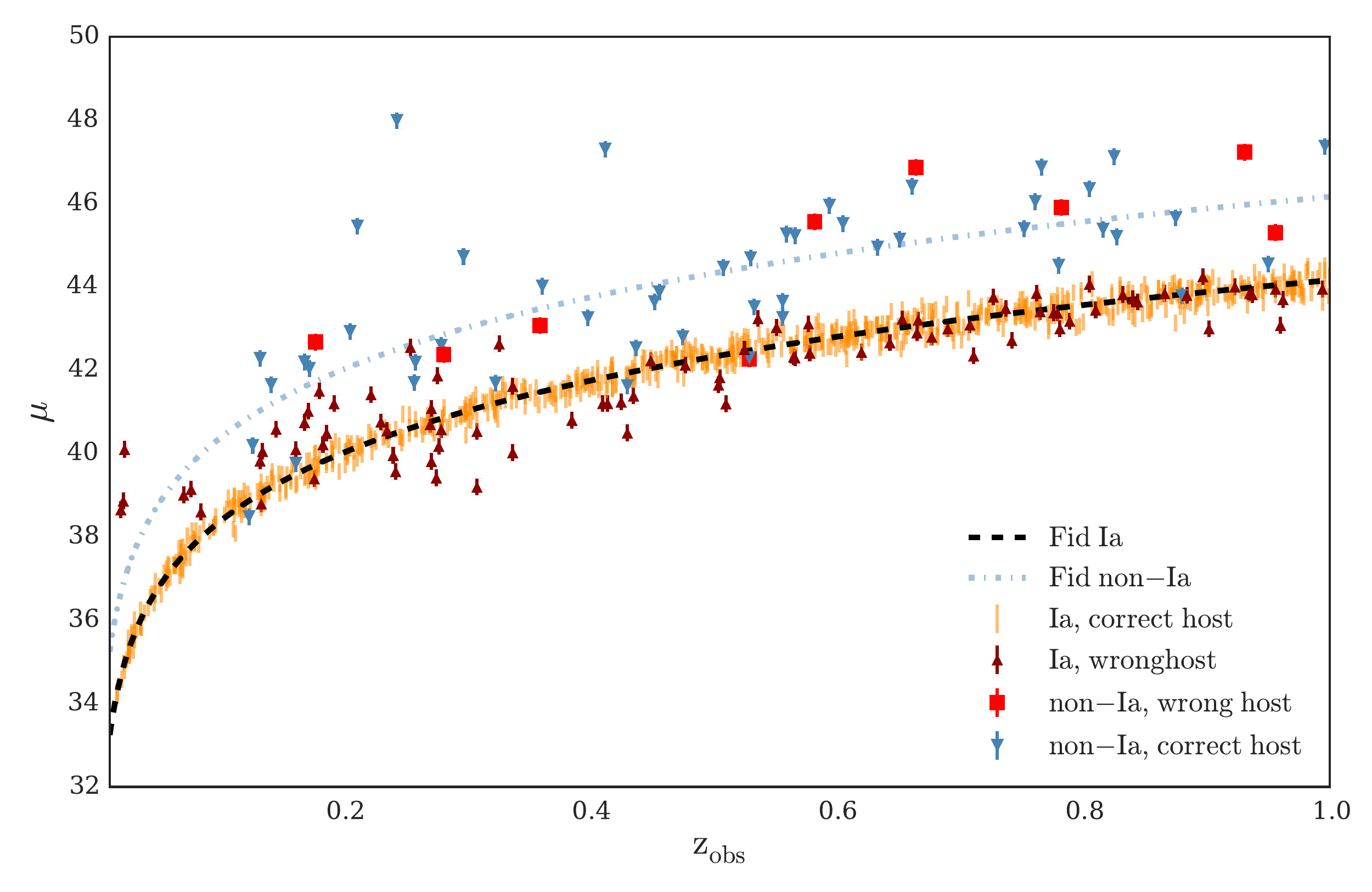}
\caption{Distance moduli for 1000 SNe with two types of contamination: $\sim 9\%$ wrong host-galaxy identification (and hence incorrect redshift, shown as red and maroon points) and $\sim 5\%$ non-Ia contaminants (shown as blue and red data points). The red data points represent SNe that have both the wrong redshift and are non-Ia. For cases with the wrong host galaxy redshift we assign a Gaussian error with a  conservative standard deviation of $0.1$. The fiducial Ia and non-Ia distance modulus are shown as the thick and thin dashed lines respectively. Figure~\ref{fig:contourspec} shows how zBEAMS is able to untangle both forms of contamination with little increase in error contour size, while applying the standard MCMC approach ignoring the contamination leads to very significant biases.}
\label{fig:spec_Hubble}
\end{figure}

\begin{figure}[tbp]
\centering
\includegraphics[width=14cm]{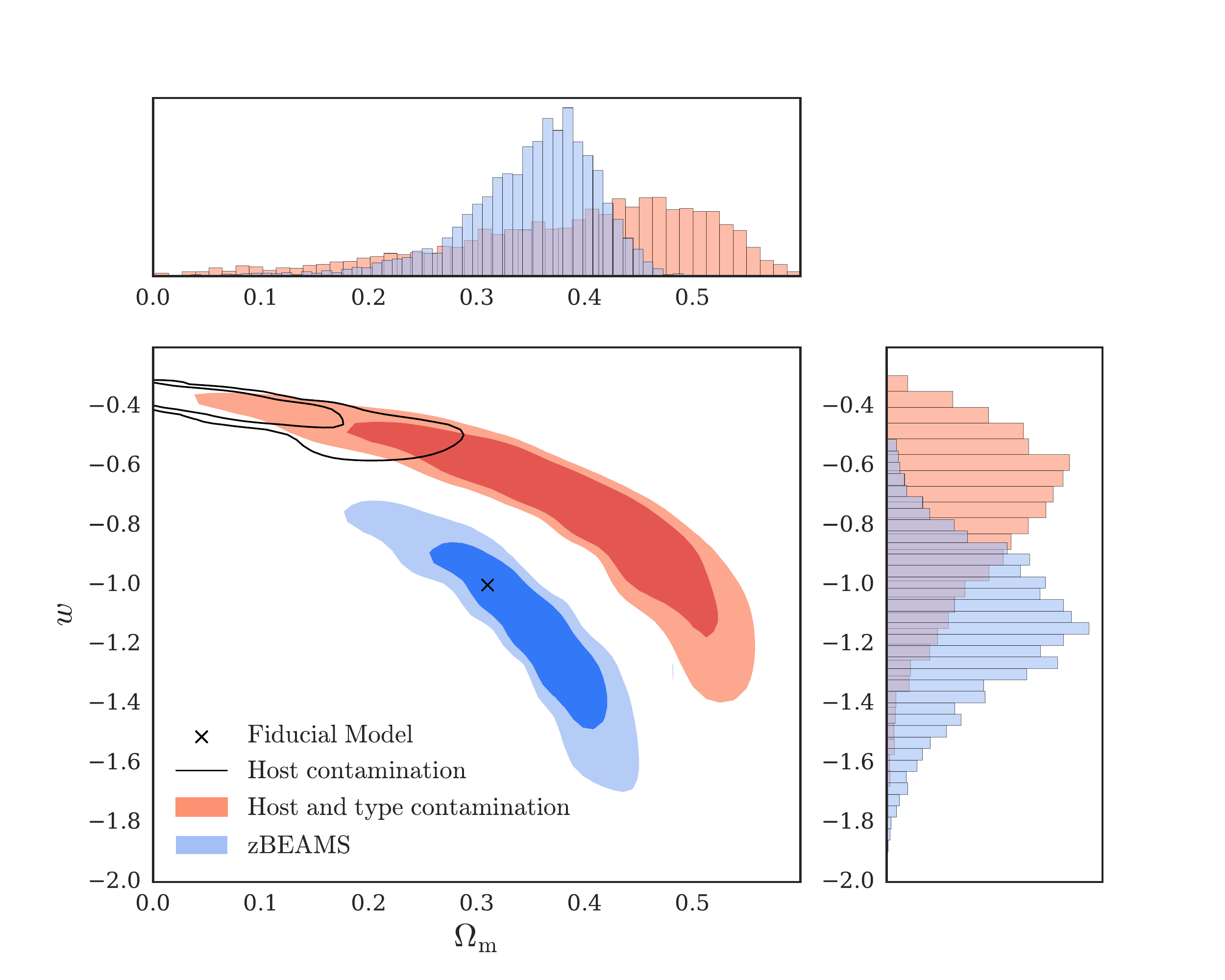}
\caption{Contour plots for $w$ and $\Omega_m$ showing the $68\%$ and $95\%$ credible intervals for the three instances we consider in the spectroscopic case. The black cross shows the fiducial model from which the data were generated. We show the biased posteriors for a dataset with host contamination only (black outlined contours) and for a dataset with both host and type contamination (red solid contours), where in both cases we use the standard likelihood without accounting for the host redshift and type contamination. Finally the blue solid contours show the zBEAMS posterior on the same doubly-contaminated dataset, where we account for the $9\%$ redshift contamination with Gaussian redshift confusion ($\sigma_z \sim 0.1$) and $5\%$ non-Ia type confusion. We find that zBEAMS handles both the redshift and type contamination with little increase in computational complexity or error ellipse area. Top and right panels show the 1D marginalised histograms for $\Omega_m$ and $w$ respectively for the standard likelihood (red) and the zBEAMS likelihood (blue).}
\label{fig:contourspec}
\end{figure}

The unbiased SNe data is generated with a dispersion of $0.2~\rm{mag}$. For the biased dataset, we assume the host galaxies have spectroscopically confirmed redshifts, but the supernova is observed using photometry, hence the supernova type is not known and it is not always clear which galaxy the supernova belongs to if multiple galaxies lie within a small angular distance of one another. We also considered that supernovae with $z<0.1$ will be identified spectroscopically. We assume $5\%$ type misidentification where the non-Ia population is offset from the Ia population by $2~\rm{mag}$, and has a Gaussian dispersion of $1.5~\rm{mag}$ (similar to \cite{Hlozekbeams}). A more realistic distribution can be used with the same method, provided the form is known. We assume a $9\%$ host misidentification \cite{Gupta2016}, where the misidentified host redshift is drawn from a normal distribution $z \sim \mathcal{N}(z_{\rm{true}}, 0.1^2)$. 

In figure~\ref{fig:spec_Hubble} we plot the distance modulus of a contaminated SNe Ia dataset. The biased dataset we analyse using both the standard likelihood (which does not take redshift error into account) and with the zBEAMS likelihood. We use the zBEAMS posterior in eq.~\eqref{eq:fullposterior} to fully marginalise over both type and redshift uncertainties and thus produce unbiased cosmological estimates. In this analysis, we solve for $\Omega_m$, $H_0$ and $w$ while we assume the parameters of the populations (such as the magnitude offset and standard deviation of the non-Ia population) are known. However, it would be simple to solve for these simultaneously as done in earlier BEAMS papers \cite{Hlozekbeams,Knights2012}. We infer the marginalised posterior distribution for $w$ and $\Omega_m$ for each of these three instances, and their respective contours are shown in figure~\ref{fig:contourspec}.

\subsection{Photometric Case}

\begin{figure}[tbp]
\includegraphics[width=15cm]{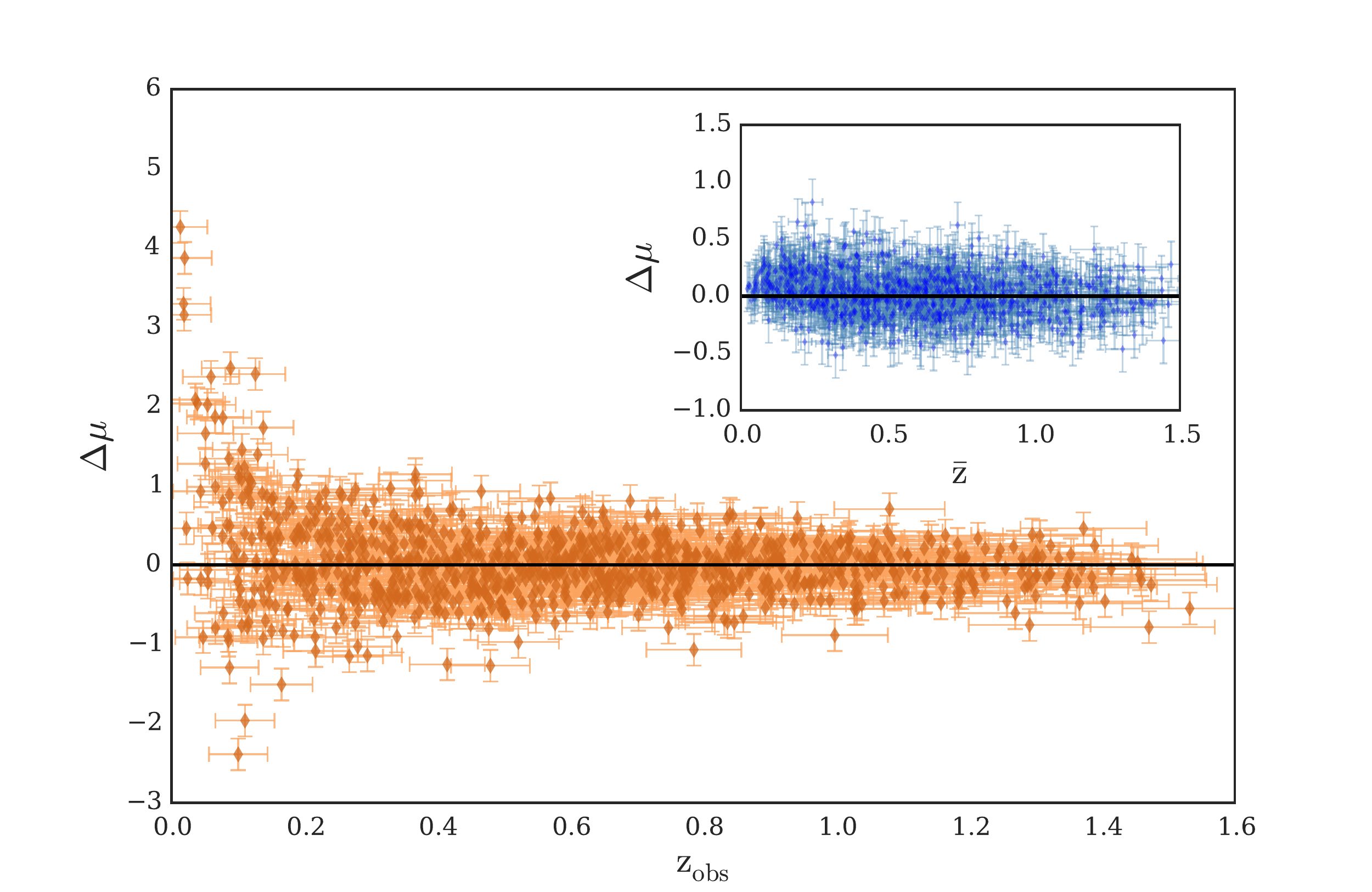}
\centering
\caption{Photometric Hubble residuals for 998 SNIa drawn from the redshift prior distribution $P(z) \sim z e^{-3z}$ with photometric redshift errors drawn from a Gaussian with $\sigma_z = 0.04(1+z)$. Main figure: residuals plotted against $\z$ (gold). Inset: residuals plotted against the  redshifts recovered from the MCMC chains, $\bar{z}$, (blue). The redshift uncertainties cause a large fraction of the data to appear more than $3\sigma$ away from the fiducial model, with some points more than $20\sigma$ away (the error on $\mu$ is $0.2$ mag in all cases). Instead zBEAMS handles these large excursions by effectively putting the supernovae at the correct redshift (as shown in the inset where the majority of points are less than $2\sigma$ from the fiducial model) and recovers the unbiased cosmology contours shown in figure~\ref{fig:contourphot} without significant decrease in precision.}
\label{fig:hubble_residuals_phot}
\end{figure}

Here we consider the case where the redshift of the host galaxy is obtained photometrically. Now to use zBEAMS, the marginalisation over redshift must be performed numerically, as the integral in eq.~\eqref{eq:anights} has no analytic solution. We assume for this work that the redshift uncertainties are Gaussian distributed with a standard deviation of $0.04(1+z)$, though any more realistic distribution can be assumed with little change in complexity. We simulate 998 SNe from a redshift distribution given by $P(z) \sim z e^{-\beta z}$ spanning the redshift range $z=0.015-1.4$, with $\beta=3$. Note that this distribution is also the prior on the fitted redshifts and in reality one would need to fit for $\beta$ simultaneously with all other parameters. While we do not address this here, this distribution could be extended to include modeling of instrumental selection effects, in addition to intrinsic supernova rate information. As before, we assumed Gaussian errors with dispersion $0.2~\rm{mag}$ in the distance modulus. In figure~\ref{fig:hubble_residuals_phot} one can see the magnitude residuals for the observational redshift (main figure) and for the redshifts recovered by the zBEAMS analysis (inset figure). 

\begin{figure}[tbp]
\centering
\includegraphics[width=14cm]{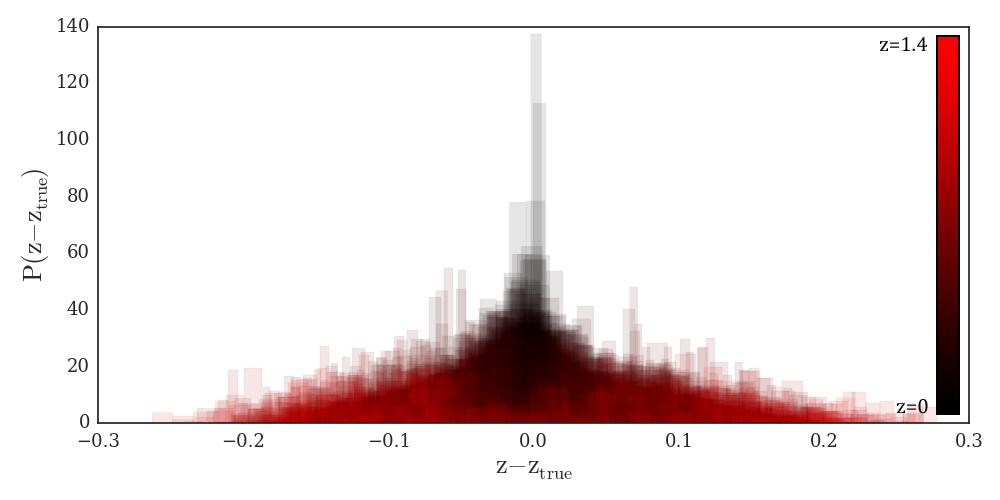}
\caption{Stacked one-dimensional histograms for all 998 redshifts from the zBEAMS analysis of the data in figure~\ref{fig:hubble_residuals_phot}. For each supernova we plot the histogram relative to the true redshift, demonstrating that zBEAMS recovers, on average, the true redshift for each supernova. Each histogram is coloured by its redshift: black corresponding to low redshifts and red corresponding to high redshifts, showing that the recovered redshifts are less precise for increasing redshift, as expected due to the $(1+z)$ scaling of the photometric redshift error and the flattening of the Hubble diagram.}
\label{fig:phot_hubble2}
\end{figure}

\begin{figure}[tbp]
\centering
\includegraphics[width=14cm]{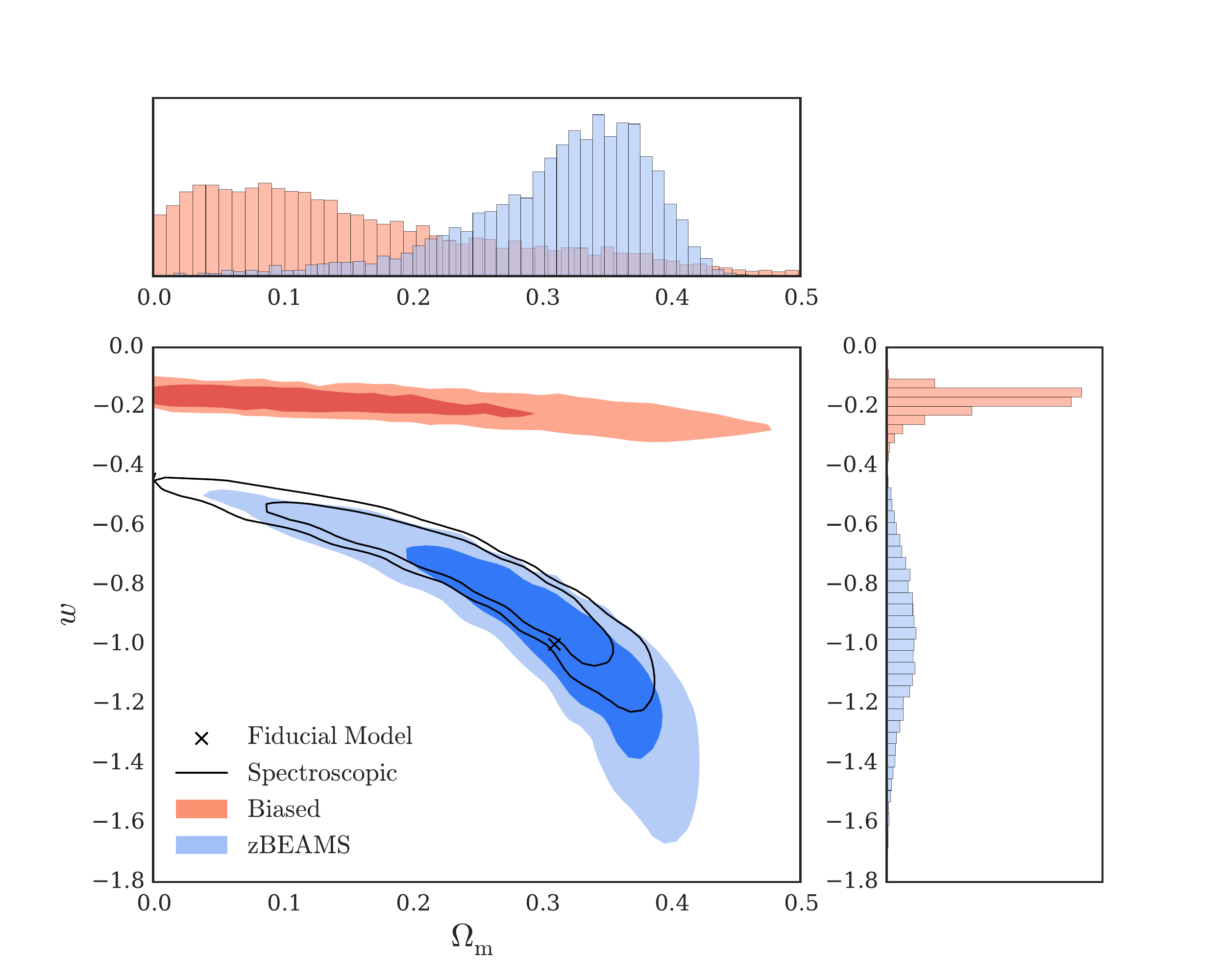}
\caption{Contour plots for $w$ and $\Omega_m$ showing the $68\%$ and $95\%$ credible intervals for the three instances we consider in the photometric case with Gaussian redshift errors $\sigma_z = 0.04(1+z)$. 998 SNIa without type contamination were simulated with redshifts drawn from a prior distribution given by $P(z) \sim z e^{-3z}$. The black cross shows the fiducial model from which the data was generated. The black outlined contours show the ideal ``model answer'' posterior where we use the standard likelihood on the unbiased/spectroscopic dataset. The red solid contours illustrate the posterior for the standard MCMC on the biased dataset (without taking into account the redshift errors). The blue solid contours show the posterior found using zBEAMS on the same biased dataset. Top and right panels show the 1D marginalised histograms for $\Omega_m$ and $w$ respectively for the standard likelihood (red) and the zBEAMS likelihood (blue).}
\label{fig:contourphot}
\end{figure}

We then do a MCMC using the standard likelihood (where we only solve for 3 cosmological parameters assuming incorrectly that all the redshifts are correct). The result is plotted in figure \ref{fig:contourphot} in shades of red. Note that the result is clearly biased with respect to the solid black contours which are obtained using the true SNe redshifts. The blue contours in figure \ref{fig:contourphot} show the result when applying zBEAMS to the biased dataset. We used a block Metropolis-Hastings sampling method -- affectionately dubbed ``Arabian nights” -- to fit for 1001 parameters simultaneously (3 cosmological parameters and 998 redshifts), i.e., numerically computing the posterior given by eq.~\eqref{eq:anights}. The block Metropolis-Hastings proceeds identically to the usual Metropolis-Hastings sampling algorithm, except that parameters are updated in blocks instead of updating all parameters every step. We took the 3 cosmological parameters and each redshift as a block, but experimented with block sizes of 1-10 redshifts. 

The block size has little impact on accuracy, as long as the blocks are small enough not to reduce the acceptance ratio, they do however impact on algorithm speed. We assume that each supernova redshift has a prior coming from the host galaxy (or from the supernova lightcurve itself) which we take to be Gaussian centred on the observed redshift with standard deviation of $0.04(1+\z)$. The prior on the overall SNIa redshift distribution was taken to be $P(z) = z e^{-\beta z}$, where we fixed the value of $\beta$ to 3. In a case with real data, one would need to fit for these hyperparameters as well.

We found that the block Metropolis-Hastings recovers the true redshifts of the low-z supernovae ($\sigma_z = 0.02$ for $z < 0.25$) well, with worsening performance as the redshift increases. This is due to two effects: first we assume the photometric redshift error scales with $(1+z)$ and secondly the Hubble diagram progressively flattens out at $z > 0.25$ removing the signal that allows MCMC to constrain the redshift. This can be clearly seen in figure~\ref{fig:phot_hubble2} where we show the stacked 1D histograms $z_{i,\rm chain}-z_{i,\rm true}$ for all 998 SNIa.  It can be seen that while the error increases with redshift, the redshift estimates show no systematic bias. The marginalised posterior distributions for $w$ and $\Omega_m$ for each of these instances are represented in the contour plots shown in figure~\ref{fig:contourphot}. We can see that zBEAMS recovers the correct cosmology, and contours, as desired. 

Examining figure \ref{fig:hubble_residuals_phot} we can see the origin of the bias of the standard likelihood. A large number of the data points are more than $3\sigma$ away from the model and some are over $20\sigma$ away. This is an artefact of using the wrong redshifts. The inset shows the same residual Hubble diagram when the data is instead plotted using the mean redshifts recovered from the MCMC chain for each redshift. Very few datapoints are now more than $2\sigma$ from the fiducial model, even at low redshifts where the excursions were the strongest.

This allows zBEAMS to produce unbiased cosmology contours that almost match the contour sizes of the perfect, spectroscopic case. A more standard approach to this same problem might be to significantly increase all the $\mu$-error bars of the points to account for the redshift uncertainties. We find that doing so yields biased results for reasons discussed in section~\ref{sec:photoz}. Increasing the error bars further might unbias the contours but only at the expense of significantly inflating the associated contours. 

It should be noted that accurate sampling in realistic scenarios will not be trivial since we are fitting for a posterior that typically has more unknown parameters than data points. We have explored both block Metropolis-Hastings and other algorithms such as Diffusive Nested Sampling \cite{brewer2016} as viable solutions. Hamiltonian Monte Carlo \cite{neal2012} may also be well-suited to the high-dimensionality of this problem. 

The code used in this section is available on Github at: \\
https://github.com/MichelleLochner/zBEAMS. 

\section{Conclusions and Future Work}

Future large surveys such as LSST will likely deliver large numbers ($> 10^5$) of good supernova candidates without the spectroscopic confirmation that has historically been required to use them for cosmology. This means we will have to do a cosmological analysis with a sample for which both the true type and redshift of the supernovae are unknown. Instead, only the probability distributions of both supernova type and redshift will be available.  In particular, the redshifts will be uncertain either because the redshift is only known photometrically or because the true identity of the galaxy hosting the supernova is unsure, even if the redshifts of potential host galaxies are perfectly known. 

In this paper we have shown how to achieve unbiased cosmology with such a sample, simultaneously handling both the non-Ia contamination and the problem of imprecise supernova redshifts in a unified framework. Our formalism - zBEAMS - generalises the original BEAMS formalism \cite{Kunz2007} to handle the redshift uncertainties of the supernovae by employing a hierarchical Bayesian approach. We introduce nuisance parameters for the type and redshift of each supernova in the dataset and then marginalise over these nuisance parameters using numerical sampling. In the special case where the supernova belongs to one of several galaxies each with spectroscopically known redshift, the final posterior is a simple weighted mixture model over the posteriors assuming the supernova is in each of the different potential host galaxies (see eq.~\eqref{eq:spectroz}). 

We show in figure~\ref{fig:contourspec} that a model with a $9\%$ host misidentification error leads to large bias ($\sim 3\sigma$) using the standard inference approach while zBEAMS removes the bias at essentially no extra computational cost. We also consider the case of photometric uncertainties, where we numerically marginalise over 998 redshift parameters to produce the unbiased contours in figure~\ref{fig:contourphot}.   

There are a number of ways in which this work can be extended in a straight-forward way: 
\begin{itemize}

\item In this analysis we have assumed that the probability of belonging to a given host galaxy $\gamma$, encoded in the terms $P(z | \gamma)$, and the probability of being a given type of supernova, $\tau$, encoded in $P(\tau | z)$, are known {\em apriori}. It would be interesting to extend our formalism to allow these to be partially known nuisance parameters that are estimated by available data. 

\item The zBEAMS formalism could be extended to include correlations with host galaxy information, such as host influence on Hubble residuals via stellar mass etc. \cite{hostgaleffects1, hostgaleffects2}. 

\item We found that the redshift distribution plays an important role for the photometric redshift case. A much more complex model could be used than the one we assumed which could include some systematic effects and allow one to learn something about supernova rates (See Malz \& Peters, {\em et al.} (in prep.)  for upcoming work on this problem in the LSST context).

\item While we have presented the zBEAMS formalism emphasising its generic nature for any data $D$,  in our examples we took $D$ to be the measured distance moduli. It would be useful to develop zBEAMS specifically for the case in which $D$ is the set of lightcurve flux measurements in different bands; i.e. one step further back in the analysis chain.  

\item As discussed in detail in (\ref{malmquist}), realistic supernova surveys censor the true SN population because of the magnitude limits of the telescope and cuts performed during the analysis. Selection effects within a Bayesian framework have already been extensively covered in e.g. \citep{Rubin2015} and could be incorporated into the zBEAMS likelihood.

\end{itemize}
These extensions are left to future work.

\acknowledgments
The authors thank Alan Heavens, Nosiphiwo Zwane, Laura Richter and other participants of the 2016 Bayes School in Stellenbosch for early contributions and Rahul Biswas, Rick Kessler, Bob Nichol and Hiranya Peiris for comments on the draft. We also thank the anonymous referee for comments. BB thanks Ren\'ee Hlozek and Martin Kunz for long-term collaborations related to BEAMS and Roberto Trotta for discussions. ML thanks Brendon Brewer for help with DNest. JF thanks M\'ario G. Santos for useful discussions and Marta Spinelli for help with MCMC techniques. This paper originated at the 2016 Bayes School and Workshop in November 2016 funded by NITHeP, and grew out of Alan Heaven's lectures at the same School. The authors acknowledge support from the NRF, the South African Square Kilometre Array Project, the National Astrophysics and Space Science Programme and AIMS. We acknowledge use of the Hubble ESA Archive for images. This paper and the research it describes were created using scrum methodology.

\bibliographystyle{plain}
\bibliography{jcap_zBEAMS}

\providecommand{\href}[2]{#2}\begingroup\raggedright\begin{thebibliography}{10}

\def\physrep{Physics Reports}
\def\apj{Astrophysical Journal}
\def\mnras{MNRAS}
\def\jcap{JCAP}
\def\aap{Astronomy and Astrophysics}
\def\aj{Astronomical Journal}
\def\prd{Physical Review D}
\def\pasp{Publications of the ASP}

\bibitem{BAO1}
D.~H. Weinberg et al., \emph{Observational probes of cosmic acceleration},
\emph{\physrep} {\bf 530} (2013) 87-255
arXiv:1201.2434

\bibitem{BAO2}
B.~A. Bassett and R. Hlozek, \emph{Baryon acoustic oscillations},
\emph{Dark Energy: Observational and Theoretical Approaches} P. Ruiz-Lapuente (2010) 246

\bibitem{LSST2009}
LSST Science Collaboration and P.~A. Abell, \emph{LSST Science Book, Version 2.0},
\emph{ArXiv e-prints} (2009) 
arXiv:0912.0201

\bibitem{Mandel2009}
K.~S. Mandel et al., \emph{Type Ia Supernova Light-Curve Inference: Hierarchical Bayesian Analysis in the Near-Infrared},
\emph{\apj} {\bf 704} (2009) 629-651
arXiv:0908.0536

\bibitem{Rubin2015}
D. Rubin et al., \emph{UNITY: Confronting Supernova Cosmology's Statistical and Systematic Uncertainties in a Unified Bayesian Framework},
\emph{\apj} {\bf 813} (2015) 137
arXiv:1507.01602

\bibitem{kesslerBEAMS}
R. Kessler and D. Scolnic, \emph{Correcting Type Ia Supernova Distances for Selection Biases and Contamination in Photometrically Identified Samples},
\emph{\apj} {\bf 836} (2017) 56
arXiv:1610.04677

\bibitem{BAHAMAS}
H. Shariff, \emph{BAHAMAS: New Analysis of Type Ia Supernovae Reveals Inconsistencies with Standard Cosmology},
\emph{\apj} {\bf 827} (2016) 1
arXiv:1510.05954

\bibitem{Ma}
C. Ma et al., \emph{Application of Bayesian graphs to SN Ia data analysis and compression},
\emph{\mnras} {\bf 463} (2016) 1651-1665
arXiv:1603.08519

\bibitem{Jennings}
Jennings E. et al., \emph{A new approach for obtaining cosmological constraints from Type Ia Supernovae using Approximate Bayesian Computation},
\emph{ArXiv e-prints} (2016)
arXiv:1611.03087

\bibitem{Betoule2014}
Betoule M. et al., \emph{Improved cosmological constraints from a joint analysis of the SDSS-II and SNLS supernova samples},
\emph{\aap} {\bf 568} (2014) A22
arXiv:1401.4064

\bibitem{Chambers2016}
Chambers K.~C. et al., \emph{The Pan-STARRS1 Surveys},
\emph{ArXiv e-prints} (2016)
arXiv:1612.05560

\bibitem{psnid}
M. Sako et al., \emph{Photometric Type Ia Supernova Candidates from the Three-year SDSS-II SN Survey Data},
\emph{\apj} {\bf 738} (2011) 162
arXiv:1107.5106

\bibitem{Kessler2010}
R. Kessler et al., \emph{Photometric Estimates of Redshifts and Distance Moduli for Type Ia Supernovae},
\emph{\apj} {\bf 717} (2010) 40-57
arXiv:1001.0738

\bibitem{Wang2015}
Y. Wang et al.,\emph{Analytic photometric redshift estimator for Type Ia supernovae from the Large Synoptic Survey Telescope},
\emph{\mnras} {\bf 451} (2015) 1955-1963
arXiv:1501.06839

\bibitem{Moller2016}
A. M{\"o}ller et al., \emph{Photometric classification of type Ia supernovae in the SuperNova Legacy Survey with supervised learning},
\emph{\jcap} {\bf 12} (2016) 008
arXiv:1608.05423

\bibitem{Hlozekbeams}
R. Hlozek et al., \emph{Photometric Supernova Cosmology with BEAMS and SDSS-II},
\emph{\apj} {\bf 752} (2012) 79
arXiv:1111.5328

\bibitem{Campbell}
H. Campbell  et al., \emph{Cosmology with Photometrically Classified Type Ia Supernovae from the SDSS-II Supernova Survey},
\emph{\apj} {\bf 763} (2013) 88
arXiv:1211.4480

\bibitem{Olmstead2013}
M.~D. Olmstead et al., \emph{Host Galaxy Spectra and Consequences for Supernova Typing from the SDSS SN Survey},
\emph{\aj} {\bf} (2014) 75
arXiv:1308.6818

\bibitem{Yuan}
F. Yuan et al., \emph{OzDES multifibre spectroscopy for the Dark Energy Survey: first-year operation and results},
\emph{\mnras} {\bf 452} (2015) 3047-3063
arXiv:1504.03039

\bibitem{Gupta2016}
R.~R. Gupta et al., \emph{Host Galaxy Identification for Supernova Surveys},
\emph{\aj} {\bf 152} (2016) 54
arXiv:1604.06138

\bibitem{Guy2007}
J. Guy et al., \emph{SALT2: using distant supernovae to improve the use of type Ia supernovae as distance indicators},
\emph{\aap} {\bf 466} (2007) 11-21
arXiv:astro-ph/0701828

\bibitem{Metropolis1953}
N. Metropolis et al., \emph{Equation of State Calculations by Fast Computing Machines},
\emph{The Journal of Chemical Physics} {\bf 21} (1953) 1087-1092

\bibitem{Hastings1970}
W.~K. Hastings, \emph{Monte Carlo sampling methods using Markov chains and their applications},
\emph{Biometrika} {\bf 57} (1970) 97-109

\bibitem{March2011}
M.~C. March et al., \emph{Improved constraints on cosmological parameters from Type Ia supernova data},
\emph{\mnras} {\bf 418} (2011) 2308-2329
arXiv:1102.3237

\bibitem{gull}
S.~F. Gull, \emph{Bayesian Data Analysis: Straight-line fitting},
\emph{Maximum Entropy and Bayesian Methods} Springer Netherlands (1989) 511-518

\bibitem{doublefisher}
A.~F. Heavens et al., \emph{Generalized Fisher matrices},
\emph{\mnras} {\bf 445} (2014) 1687-1693 
arXiv:1404.2854

\bibitem{Kunz2007}
M. Kunz et al., \emph{Bayesian estimation applied to multiple species},
\emph{\prd} {\bf 75} (2007) 103508
arXiv:astro-ph/0611004

\bibitem{Knights2012}
M. Knights et al., \emph{Extending BEAMS to incorporate correlated systematic uncertainties},
\emph{\jcap} {\bf 1} (2013) 039
arXiv:1205.3493

\bibitem{challenge}
R. Kessler et al., \emph{Results from the Supernova Photometric Classification Challenge},
\emph{\pasp} {\bf 122} (2010) 1415
arXiv:1008.1024

\bibitem{Sereno2016}
M. Sereno, \emph{A Bayesian approach to linear regression in astronomy},
\emph{\mnras} {\bf 455} (2016) 2149-2162
arXiv:1509.05778

\bibitem{Mackay2003}
D~.J~.C MacKay, \emph{Information Theory, Inference and Learning Algorithms},
Cambridge University Press (2003)

\bibitem{Planck2015}
Planck collaboration and P.~A.~R. Ade, \emph{Planck 2015 results. XIII. Cosmological parameters},
\emph{\aap} {\bf 594} (2016) A13
arXiv:1502.01589

\bibitem{brewer2016}
B.~J. Brewer et al., \emph{Diffusive Nested Sampling},
\emph{G. Stat Comput} {\bf} (2011) 649
arXiv:0912.2380

\bibitem{neal2012}
R.~M. Neal, \emph{MCMC using Hamiltonian dynamics},
\emph{ArXiv e-prints} (2012)
arXiv:1206.1901

\bibitem{hostgaleffects1}
M. Sullivan et al., \emph{The dependence of Type Ia Supernovae luminosities on their host galaxies},
\emph{\mnras} {\bf 406} (2010) 782-802
arXiv:1003.5119

\bibitem{hostgaleffects2}
H. Lampeitl et al., \emph{The Effect of Host Galaxies on Type Ia Supernovae in the SDSS-II Supernova Survey},
\emph{\apj} {\bf 722} (2010) 566-576
arXiv:1005.4687

\end{thebibliography}\endgroup

\end{document}